\documentclass[aps,prb,twocolumn,showpacs]{revtex4}
\usepackage{amssymb}
\usepackage[latin9]{inputenc}
\usepackage{amsbsy}
\usepackage{amstext}
\usepackage{graphicx}
\usepackage{esint}

\begin{document}

\title{Linear magnetoresistance in the charge density wave state of
quasi-two-dimensional rare-earth tritellurides}
\author{A.A.~Sinchenko$^{1,2,3}$,P.D.~Grigoriev$^{4,5,6\ast}$, P.~Lejay$^{7}$,
and P.~Monceau$^{7}$}
\date{\today}

\begin{abstract}
We report measurements of the magnetoresistance in the charge density wave
(CDW) state of rare-earth tritellurides, namely TbTe$_3$ and HoTe$_3$. The
magnetic field dependence of magnetoresistance exhibits a temperature
dependent crossover between a conventional quadratic law at high $T$ and low 
$B$ and an unusual linear dependence at low $T$ and high $B$. We present a
quite general model to explain the linear magnetoresistance taking into
account the strong scattering of quasiparticles on CDW fluctuations in the
vicinity of "hot spots" of the Fermi surface (FS) where the FS
reconstruction is the strongest.
\end{abstract}

\pacs{71.45.Lr, 72.15.Gd, 72.15.Nj}
\maketitle

\address{$^{1}$Kotel'nikov Institute of Radioengineering and Electronics of RAS,
	Mokhovaya 11-7, 125009 Moscow, Russia} 
\address{$^{2}$National Research
Nuclear University (MEPhI), 115409 Moscow,Russia} 
\address{$^{3}$M.V. Lomonosov Moscow State University, 119991 Moscow, Russia} 
\address{$^{4}$L.D. Landau Institute for Theoretical Physics, 142432 Chernogolovka,
	Russia} 
\address{$^{5}$P.N. Lebedev Physical Institute, RAS, 119991,
Moscow, Russia} 
\address{$^{6}$National University of Science and Technology "MISiS", 119049 Moscow,
	Russia} 
\address{$^{7}$Universt\'{e} Grenoble Alpes, CNRS, Grenoble INP,
Institut NEEL, 38042 Grenoble, France}

\section{Introduction}

\label{Intr}

Interaction between pairs of quasiparticles often leads to broken-symmetry
ground states in solids. Typical examples are the formation of Cooper pairs
in superconductors, or charge (CDW) and spin (SDW) density waves driven by
electron-phonon and electron-electron interactions respectively \cite%
{Gruner,Monceau12}. A CDW ground state is characterized by a spatial
modulation $\sim\cos(Qx+\varphi)$ of the electron density and a periodic
lattice distortion with the same $Q_{CDW}=2k_{F}$ wave vector inducing
opening of a gap, $\Delta$, in the electron spectrum. From one-dimensional
(1D) weak coupling mean field theories, with $\Delta/E_{F}\ll1$, the Peierls
instability is driven by the electronic energy gain which originates mostly
from the Fermi surface (FS) nesting with $Q=2k_{F}$.

In the case of not complete nesting in quasi-one-dimensional (Q1D) compounds
or in the case of quasi-two-dimensional (Q2D) or three-dimensional (3D)
conductors the ground state in the CDW state is semimetallic because
electron and hole pockets remain in the FS. Properties of these carriers can
be modified by the CDW ordering. One of the methods to understand this
possible modification is to study magnetoresistance.

In conventional metals, the Lorentz force caused by an applied magnetic
field changes the electron trajectory and gives rise to a positive
magnetoresistance (MR) which increases quadratically with the strength of
the field \cite{Abrik,Ziman,Kittel}. Only in a few cases the MR
may grow linearly with the field (LMR). For the first time such type of
behavior was observed by Kapitza \cite{Kapitza} in polycrystalline metals.
It was shown that LMR is attributed to the presence of open Fermi surfaces.
The quantum mechanism of LMR was proposed by Abrikosov \cite%
{Abrikosov98,Abrikosov99}. In his model LMR is realized basically in gapless
semiconductors or semimetals with a linear energy spectrum and with a very
small carrier concentration, so that only one Landau band participates in
the conductivity. Parish and Littlewood \cite{Littlewood03} considered a
macroscopically disordered and strongly inhomogeneous semiconductor and
showed that a classical mechanism will give LMR in this case. In Ref. [%
\onlinecite{Herring60}] it was shown that LMR may occur in weakly
inhomogeneous systems, for fields where the cyclotron orbit period exceeds
the scattering time.

From many published works one can also conclude that this unusual LMR may be
a common feature of CDW systems. Indeed, LMR was observed in Q1D compounds
exhibiting a CDW with incomplete FS nesting such as NbSe$_{3}$ \cite%
{Richard87,Coleman90} and in PdTeI \cite{Lei16}. Effect of LMR was reported
for Q2D compounds with a CDW: transition metal dichalcogenides 2H-NbSe$_2$;
2H-TaSe$_2$ \cite{Naito82}; 1T-TaTe$_{2}$ \cite{Chen17}, 1T- NbTe$_{2}$ \cite%
{Chen17Arx}, monophosphate tingsten bronzes (PO$_2$)$_4$(WO$_3$)$_{2m}$ for
m=4.6 \cite{Rotger94}; molybdenum purple bronze, K$_{0.9}$Mo$_{4}$O$_{11}$
and molybdenum oxides $\eta$-Mo$_{4}$O$_{11}$ \cite{Schlenker}. In the
present work we have studied galvanomagnetic properties in another type of
Q2D compounds with a CDW, namely, rare-earth tritellurides. We have measured
magnetoresistance in the temperatures range across the Peierls transition
temperature and show that effectively LMR appears below this temperature.

Rare-earth tritellurides $R$Te$_{3}$ ($R$ =Y, La, Ce, Nd, Sm, Gd, Tb, Ho,
Dy, Er, Tm) exhibit an incommensurate CDW through the whole $R$ series with
a wave vector $\mathbf{Q}_{CDW1}=(0,0,\sim2/7c^{*})$ with a Peierls
transition temperature above 300 K for the light atoms (La, Ce, Nd). For the
heavier $R$ (Dy, Ho, Er, Tm) a second CDW occurs with the wave vector $%
\mathbf{Q}_{CDW2}=(\sim2/7a^{*},0,0)$. The superlattice peaks measured from
X-ray diffraction are very sharp and indicate a long range 3D CDW order \cite%
{Ru08,DiMasi95,Brouet08}.

Below the Peierls transition, in all $R$Te$_{3}$ compounds, the Fermi
surface is partially gapped resulting in a metallic behavior at low
temperature. The layered $R$Te$_{3}$ compounds exhibit a large anisotropy
between the resistivity along the $b$-axis and that in the $(a,c)$ plane,
typically $\sim10^{2}$ below $T_{CDW1}$ and much higher at low temperature 
\cite{Ru06}. Because the unidirectional character of the upper CDW \cite%
{Fang07,Lavagnini10R,Yao06}, a conductivity anisotropy in the $(a,c)$ plane
arises in the CDW state as was observed experimentally and explained
theoretically in Ref. \onlinecite{sinchPRL14}. The effect of the upper CDW
on the in-plane resistivity observed in experiments is very weak, no more
than a few percents of the total resistance \cite{Ru08,sinchPRL14,Ru06}.

For our study we chose two compounds: TbTe$_{3}$ as a system with
unidirectional CDW and HoTe$_{3}$ exhibiting a bidirectional CDW. In TbTe$%
_{3}$ the CDW ordering is observed well above room temperature ($T_{CDW1}$%
=336 K). In HoTe$_{3}$ the first and the second CDW transitions take place
at $T_{CDW1}=283$ K and $T_{CDW2}=110$ K correspondingly \cite{Ru08}.

\section{Experimental}

\label{Exp}

Single crystals of TbTe$_{3}$ and HoTe$_{3}$ were grown by a self-flux
technique under purified argon atmosphere as described previously \cite%
{SinchPRB12}. Thin single-crystal samples with a square shape and with a
thickness less than 1 $\mu$m were prepared by micromechanical exfoliation of
relatively thick crystals glued on a sapphire substrate. The untwinned character of
selected crystals and the spatial arrangement of crystallographic axes were
controlled by X-ray diffraction. Room temperature resistivity of crystals was $26-28 \mu\Omega$cm for TbTe$_3$ and $12-13 \mu\Omega$cm for HoTe$_3$ that is in accordance with previously reported results \cite{Ru08,sinchPRL14}. The quality of crystals was confirmed by high value of resistance residual ratio (RRR), $R(300K)/R(4K)$: 70-90 for HoTe$_3$ and more than 100 for TbTe$_3$.

The magnetic field was applied parallel to the $b$ axis, and in-plane
magnetoresistance was recorded using the van der Pauw method, sweeping the
field between $+6.5$ and $-6.5$ T. Measurements were performed at fixed
temperature in the temperature range 350-20 K with the step $\Delta T=10$ K.

\begin{figure}[t]
	\includegraphics[width=8cm]{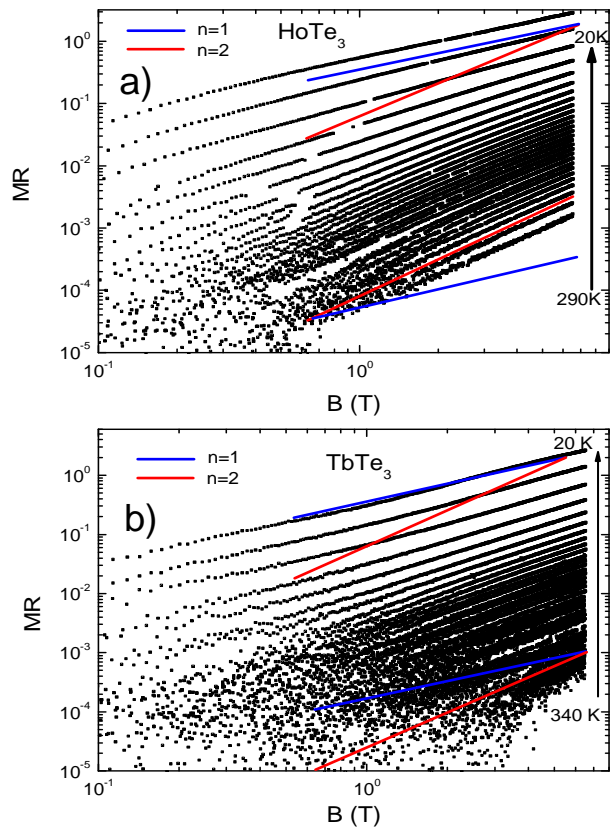}
	\caption{(color online) Magnetoresistance of HoTe$_{3}$ (a) and TbTe$_{3}$
		(b) as a function of magnetic field, $B$, in log-log scale at different
		temperatures. Blue and red straight line segments indicate linear and square
		dependencies correspondingly.}
	\label{F1}
\end{figure}

\section{Experimental results}

\label{Data}

\begin{figure}[t]
	\includegraphics[width=8cm]{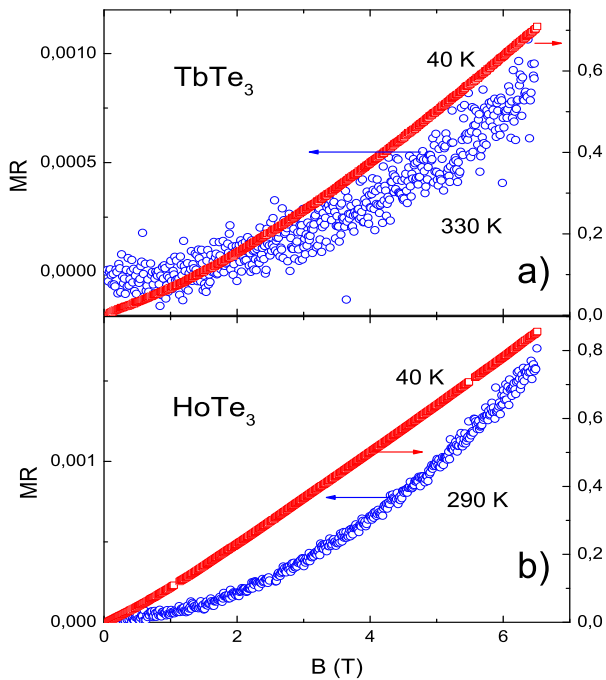}
	\caption{(color online)MR$(B)$ for TbTe$_{3}$ (a) and HoTe$_{3}$ (b),
		at temperatures above $T_{CDW}$ (open blue circles) and below $T_{CDW}$
		(open red squares).}
	\label{F2}
\end{figure}

\begin{figure}[t]
	\includegraphics[width=8.5cm]{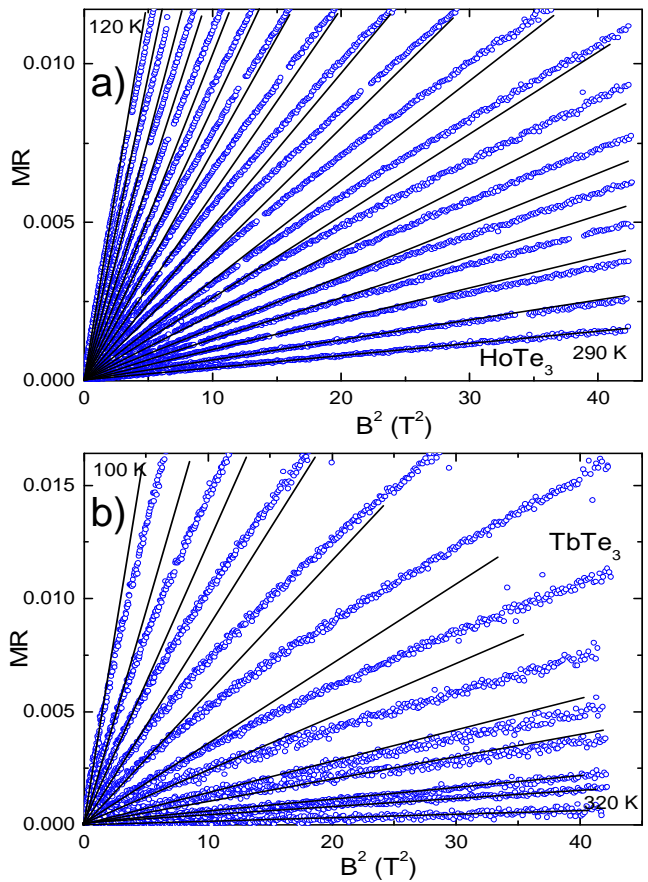}
	\caption{(color online) Magnetoresistance of HoTe$_{3}$ (a) and TbTe$_{3}$
		(b) as a function of square magnetic field, $B^{2}$, at different
		temperatures. Solid black lines demonstrate the deviation of $MR(B)$
		dependencies from a square law at some value of magnetic field, $B^{*}$.}
	\label{F3}
\end{figure}

\begin{figure}[t]
	\includegraphics[width=8.5cm]{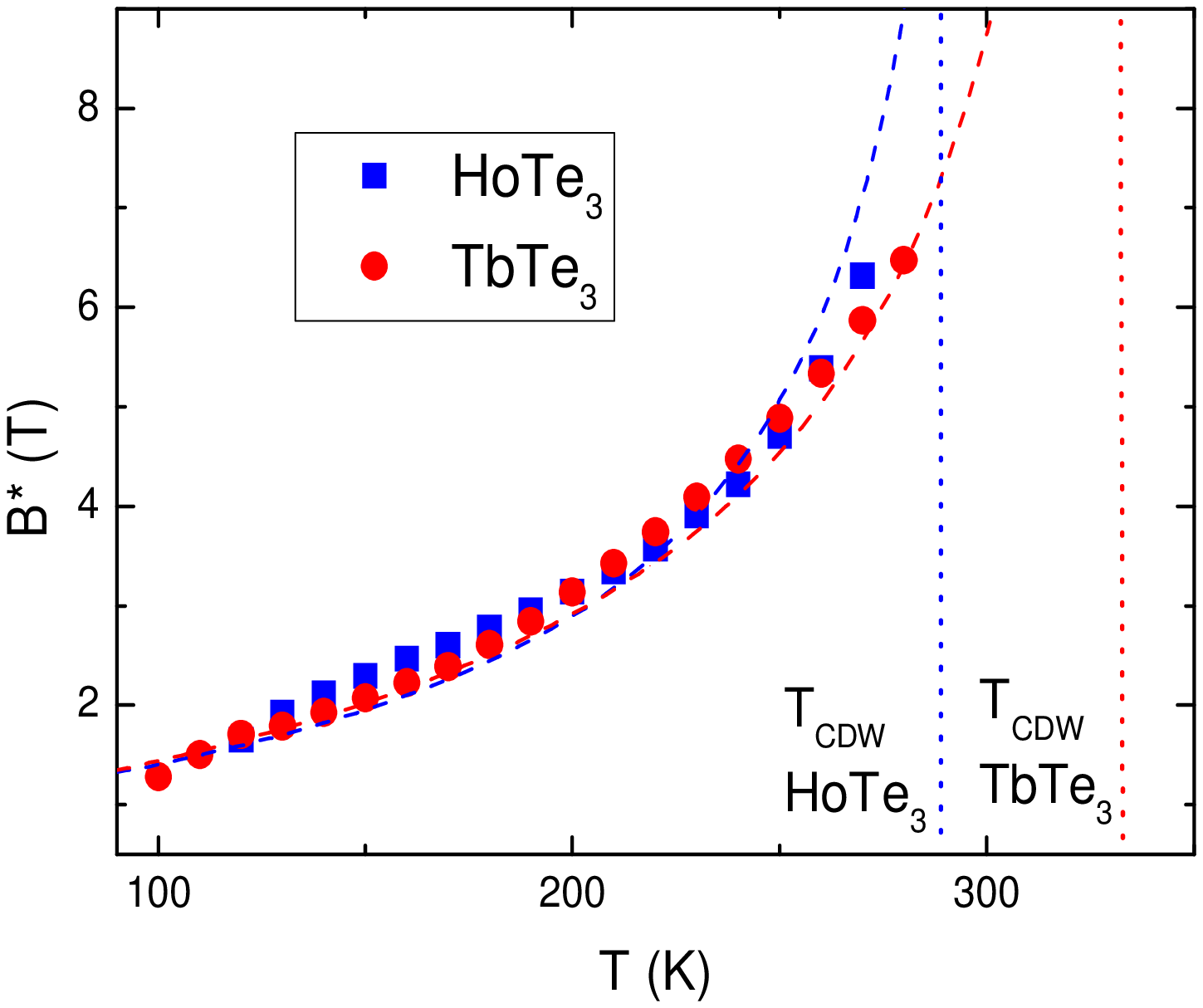}
	\caption{(color online) Temperature dependence of the characteristic field $%
		B^{*} $ for HoTe$_{3}$ (blue) and TbTe$_{3}$ (red). Dashed lines are guides
		to the eye.}
	\label{F4}
\end{figure}

The temperature variation of the field dependence of magnetoresistance,
defined as MR$=[R_{xx}(B)-R_{xx}(0)]/R_{xx}(0)$, in the temperature range
from 10 K up to the temperature well above $T_{CDW}$ is shown in Fig. \ref%
{F1} for HoTe$_{3}$ (a) and for TbTe$_{3}$ (b) in a log-log plot. Both
compounds demonstrate nearly the same behavior: magnetoresistance changes by
more than four order of magnitude as temperature $T$ decreases from 300 K to
20 K. Simultaneously, the power-law field dependence of MR changes
monotonically from quadratic (red straight line segment) at high $T$ and at
low $B$ to linear (blue straight line segment) at low $T$ and high $B$.
Note, that in the studied magnetic field range (up to 6.5 T) we never
observed any deviation of MR from quadratic law at temperatures above the
Peierls transition temperatures The examples of MR$(B)$ dependencies measured at $T$
above $T_{CDW}$ (330 K and 290 K for TbTe$_3$ and HoTe$_3$ respectively) and below $T_{CDW}$ (40 K for both compounds) are shown in Fig. \ref%
{F2}. Note, that at the same temperature $T=40$ K, the linear $R_{xx}(B)$ is
more pronounced for HoTe$_{3}$ in which two CDWs exist at this temperature.

To make this quadratic-to-linear MR crossover clearer, in Fig.\ref{F3} (a)
and (b)we plot MR as a function of square magnetic field, $B^{2}$ for HoTe$%
_{3}$ and for TbTe$_{3}$ correspondingly. Solid black lines are quadratic
dependencies which coincide with the experimental curves at low magnetic
fields. At a certain magnetic field, $B^{\ast}$, experimental dependencies
deviate from these lines. Temperature dependence of this characteristic
field $B^{\ast}$ is shown in Fig.\ref{F4}. As can be seen, $B^{\ast}$
increases rapidly or even diverges when $T$ approaches the CDW transition
temperature.

\section{Theoretical model and discussion}

Thus, most of charge-density wave systems with imperfect nesting exhibit a linear magnetoresistance that is, probably, related to the CDW electronic structure. To propose a possible explanation
of this linear MR we invoke a usually neglected scattering mechanism of
quasiparticles on the fluctuations of the order parameter of a charge
density wave, which violate the space uniformity and lead to the momentum
relaxation of quasiparticles. The scattering on CDW fluctuations is the
strongest near the so-called \textquotedblright hot spots\textquotedblright\
on the Fermi surface (FS). Somewhat similar mechanism of linear MR but above
the CDW transition temperature was proposed in Refs. \cite%
{Young1968,FalikovLinearMR}. In Ref. \cite{Young1968} the scattering in the
hot spots, with large momentum and low energy transfer, involves umklapp
processes. In Ref. \cite{FalikovLinearMR} it involves the scattering by soft
phonons, appearing due to the proximity to Peierls instability. In our case
these hot spots are the FS areas where the FS reconstruction due to CDW is
the strongest. Usually the hot spots are the ends of the ungapped FS parts.
The electron dispersion in such hot spots depends strongly on the CDW
structure, and the electrons in such hot spots may be easily scattered by
CDW fluctuations. Thus, in cuprate high-Tc superconductors such hot spots
are the ends of Fermi arcs, but the FS reconstruction is driven by the
pseudogap or antiferromagnetic ordering, rather than by CDW. In organic
metals, e.g. $\alpha $--(BEDT-TTF)$_{2}$KHg(SCN)$_{4}$, where the CDW leads
to the FS reconstruction and changes its topology,\cite{KartsLTP2014} such
hot spots are the points of intersection of the original FS and the FS
shifted by the CDW wave vector. In these hot spots the electron dispersion
changes strongly, somewhat similar to the change of electron dispersion at
the boundaries of the Brillouin zone in the weak-coupling approximation,\cite%
{Abrik} where the energy gap is formed due to periodic potential, which is
the CDW in our case. Since the periodic potential and the formed energy gap
in the electron spectrum is of the order of CDW order parameter and much
less that the Fermi energy, in high magnetic field the electron trajectories
in these hot spots are subject to magnetic breakdown in addition to the
direct scattering by CDW fluctuations. This leads to an additional indirect
scattering mechanism of conducting electrons, which may be rather strong 
\cite{KartsLTP2014}.

The electron scattering in the hot spots leads to the linear field
dependence of the scattering rate and, hence, to the linear
magnetoresistance. To show this linear field dependence of the electron
mean-free time $\tau $ we assume that in each hot spot an electron is
scattered with some probability $w_{hs}<1$; the possible origin of this
scattering is discussed later. If this hot-spot scattering is the main
scattering mechanism of conducting electrons, the corresponding electron
mean free time $\tau _{hs}=t_{hs}/w_{hs}$, where $t_{hs}$ is the mean time
between electron passing through these hot spots. This time $t_{hs}$ is
determined by the FS details \cite{Abrik}. In magnetic field $\boldsymbol{H}$
electrons move in momentum space along the Fermi surface due to the Lorentz
force, $dp/dt=(e/c)[\boldsymbol{v_{\boldsymbol{\bot }}}\times H]$, and
periodically pass through such hot spots. Hence, the mean free\ time $\tau
_{hs}$ is proportional to the length of the Fermi-surface between hot spots
divided by magnetic field $H$ strength and by electron velocity in real
space $v_{\bot }$:\cite{Abrik} 
\begin{equation}
\tau _{hs}=(c/eHw_{hs})\int dl/\boldsymbol{v_{\bot }}.  \label{tauhs}
\end{equation}%
If the electron trajectory in magnetic field is closed, its motion is
periodic given by the cyclotron (or Larmor) period $T_{L}=2\pi /\omega _{c}$%
, where the cyclotron frequency $\omega _{c}=e\hbar H/m^{\ast }c$, and $%
m^{\ast }$\ is the electron effective mass. Then $\tau _{hs}\eqsim
T_{L}/w_{hs}n_{hs}$, where $n_{hs}$ is a number of hot spots along the
cyclotron period. If the electron trajectory in magnetic field is open, it
also periodically pass through hot spots, and the length of the
Fermi-surface between hot spots is approximately given by the length $2\pi
/a^{\ast }$ of the first Brillouin zone divided by the number of hot spots
on this open trajectory, where $a^{\ast }$ is the lattice constant. Then,
according to Eq. (\ref{tauhs}), $\tau _{hs}\sim (c/eHw_{hs})2\pi /a^{\ast
}\left\vert \boldsymbol{v_{\bot }}\right\vert n_{hs}$. We see that both for
closed and open electron trajectories the hot-spot mean free time is
inversely proportional to magnetic field: $\tau _{hs}\propto 1/H$.

In the $\tau $-approximation for isotropic in-plane dispersion the
conductivity tensor $\sigma $ in magnetic field $\boldsymbol{H}$ is given by
the well-known formula\cite{Abrik} 
\begin{equation}
\sigma =\frac{n_{e}e^{2}}{m^{\ast }\left( \omega _{c}^{2}+1/\tau ^{2}\right) 
}\left( 
\begin{tabular}{ll}
$\tau ^{-1}$ & $\omega _{c}$ \\ 
$-\omega _{c}$ & $\tau ^{-1}$%
\end{tabular}%
\right) ,  \label{sDrude}
\end{equation}%
which gives for the resistivity tensor 
\begin{equation}
R=\sigma ^{-1}=\frac{m^{\ast }}{n_{e}e^{2}}\left( 
\begin{tabular}{ll}
$\tau ^{-1}$ & $-\omega _{c}$ \\ 
$\omega _{c}$ & $\tau ^{-1}$%
\end{tabular}%
\right) .  \label{RDrude}
\end{equation}%
When $\tau $ is independent of magnetic field, as in the simplest models of
electron scattering by impurities or by phonons, Eq. (\ref{RDrude}) predicts
no magnetoresistance: $\Delta R_{xx}(H)\equiv R_{xx}(H)-R_{xx}(0)=0$. The
absence of magnetoresistance in this model is the result of the Hall
electric field, which balances the Lorentz force. This balance can be
maintained and leads to zero magnetoresistance only if the drift velocity $v$%
, included in the equations of motion, is the same for all charge carriers.
Therefore in metals with several types of charge carrier, e.g. electrons or
holes from different Fermi-surface parts, the quadratic magnetoresistance
appears, which saturates at high magnetic field. Thus, for the simplest
isotropic model of only two types of carriers the calculation based on the
kinetic equation gives (see Eq. 7.163 of \cite{Ziman}) 
\begin{equation}
\frac{\Delta R_{xx}(H)}{R_{xx}(0)}=\frac{\sigma _{1}\sigma _{2}\left( \omega
_{c1}\tau _{1}-\omega _{c2}\tau _{2}\right) ^{2}}{\left( \sigma _{1}+\sigma
_{2}\right) ^{2}+\left( \omega _{c1}\tau _{1}\sigma _{1}+\omega _{c2}\tau
_{2}\sigma _{2}\right) ^{2}},
\end{equation}%
where the subscripts $1$ and $2$ denote the charge carriers of the first and
second type respectively. The quadratic dependence here comes from $\omega
_{c}$ in numerator, and the saturation at $\omega _{c}\tau \gtrsim 1$ comes
from $\omega _{c}$ in denominator. Usually, the relaxation time depends on
the speed $v_{i}$ of an individual charge carrier, so that one cannot
describe the motion of the carriers in terms of a single drift velocity even
for metals with a single electron band. Therefore,\cite{Kittel} even in
single-band metals in weak field one observes the quadratic
magnetoresistance 
\begin{equation}
\frac{\Delta R_{xx}(H)}{R_{xx}(0)}\equiv \frac{R_{xx}(H)-R_{xx}(0)}{R_{xx}(0)%
}=\frac{\alpha \left( \omega _{c}\tau \right) ^{2}}{1+\beta \left( \omega
_{c}\tau \right) ^{2}},  \label{Rxx}
\end{equation}%
where the coefficients $\alpha \sim \beta \sim 1$, which saturates in strong
field when $\omega _{c}\tau \gtrsim 1$. This quadratic magnetoresistance in
weak fields can be understood also in terms of the curvature of electron
trajectories, which geometrically reduces the electron mean-free path $%
l_{i}=\tau v_{i}$ by the quantity $\sim l_{i}\left( l_{i}/r_{L}\right) ^{2}$%
, where $r_{L}\gg l_{i}$ is the Larmor radius.\cite{Abrik} Since $%
l_{i}\propto \tau $, this effect can be taken into account in Eq. (\ref%
{RDrude}) by the renormalization of the electron mean-free time $\tau \left(
H\right) $ in weak magnetic field $H$ according to\cite{Abrik} 
\begin{equation}
\frac{\tau \left( 0\right) }{\tau \left( H\right) }=1+\frac{\alpha \left(
\omega _{c}\tau \right) ^{2}}{1+\beta \left( \omega _{c}\tau \right) ^{2}},
\label{tauH}
\end{equation}%
which leads to Eq. (\ref{Rxx}).

\begin{figure}[t]
	\includegraphics[width=8cm]{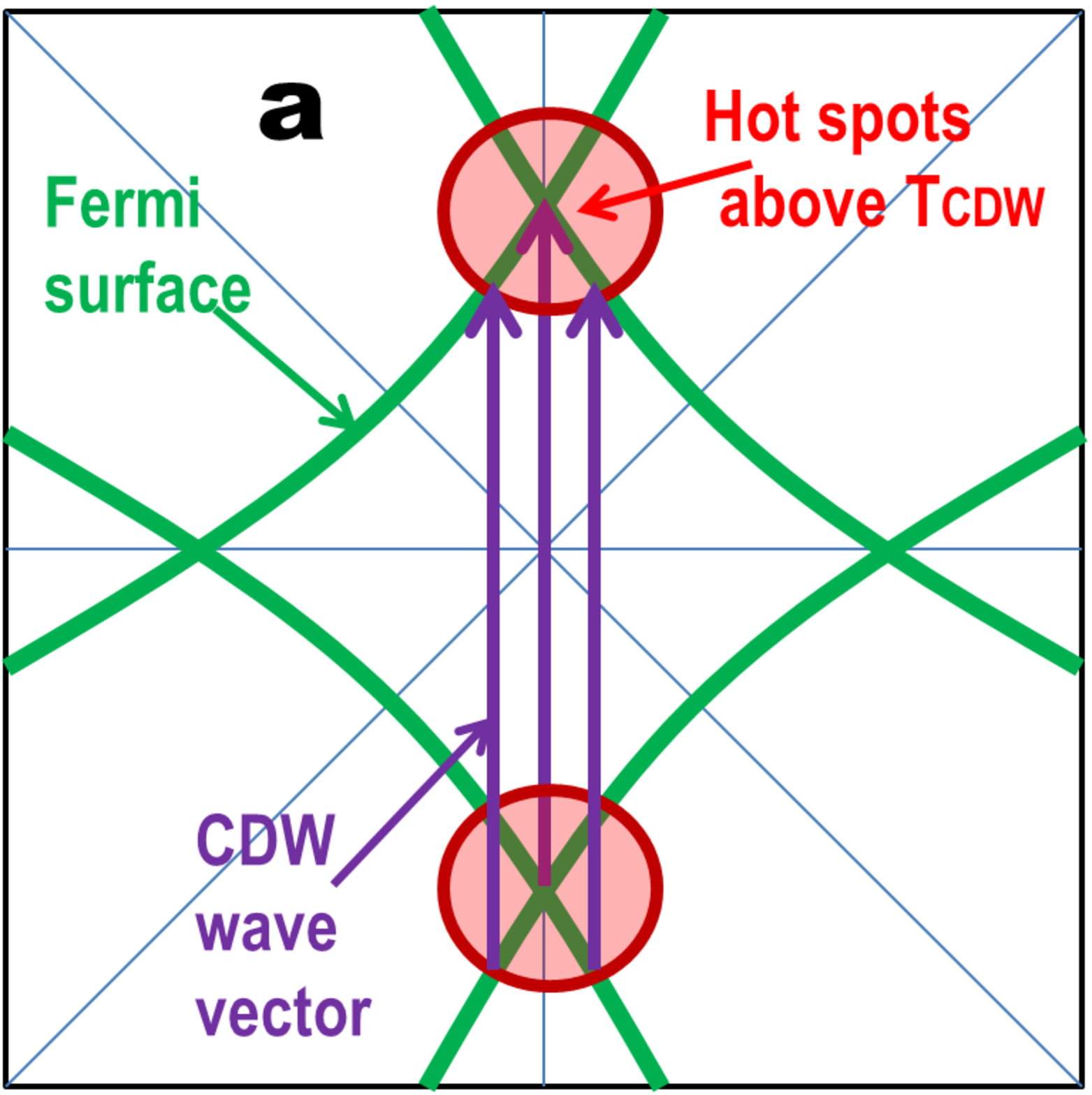} \newline
	\includegraphics[width=8.2cm]{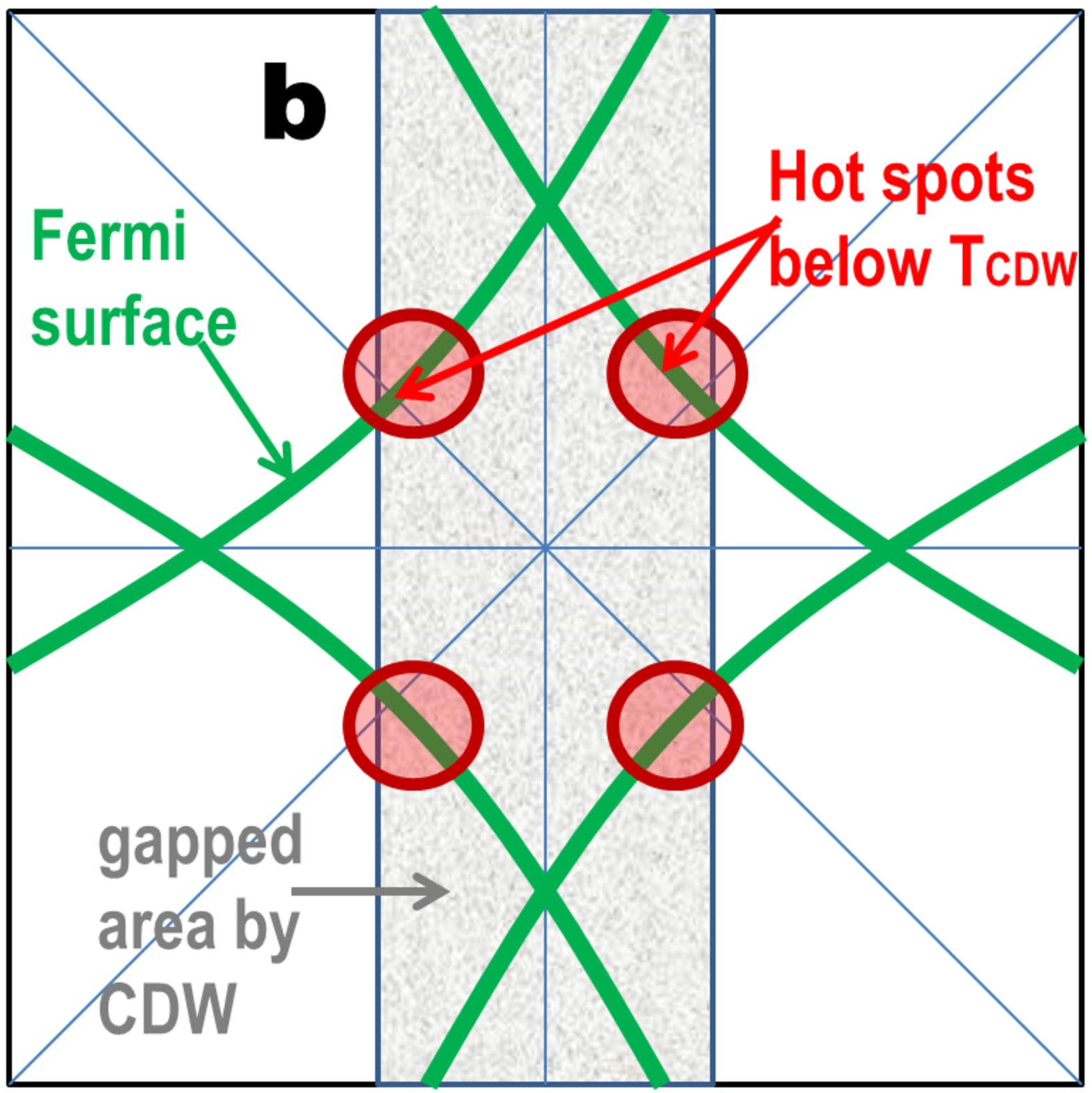}
	\caption{(color online) The position of hot spots on the Fermi surface of
		RTe $_{3}$ above, as proposed in Ref. \onlinecite{FalikovLinearMR} (a) and
		below (b) the CDW transition temperature (our work).}
	\label{F5}
\end{figure}

If the electron scattering is dominated by the scattering in the hot spots,
instead of Eq. (\ref{tauH}) we have $\tau \left( H\right) \approx \tau
_{hs}\propto 1/H$, and Eq. (\ref{RDrude}) gives $R_{xx}\propto H$. However,
in real compounds the total scattering rate $\tau ^{-1}$ is a sum of the
contribution from various mechanisms, including those from the scattering by
impurities $\tau _{i}^{-1}$ and by phonons $\tau _{ph}^{-1}$. The scattering
by phonons at high temperature is somewhat similar to scattering by
short-range impurities, as in both cases the momentum transfer during each
scattering is large and comparable to the Fermi momentum. In rather weak
magnetic field, when the Landau levels are not separated and the magnetic
quantum oscillation can be neglected,\cite{Comment1} the scattering rates $\tau _{i}^{-1}$
and $\tau _{ph}^{-1}$ depend on magnetic field according to Eq. (\ref{tauH}).
 Then $\tau ^{-1}\approx \tau _{i}^{-1}+\tau _{ph}^{-1}+\tau
_{hs}^{-1}$, and the linear MR appears only in rather strong magnetic field,
when $\tau _{hs}^{-1}>\tau _{i}^{-1}+\tau _{ph}^{-1}\equiv \tau _{i+ph}^{-1}$%
, i.e. when $\omega _{c}\tau _{i+ph}\gtrsim 2\pi /w_{hs}n_{hs}$, or when the
 quadratic magnetoresistance saturates. At high
temperature, when the scattering rate by phonons $\tau _{ph}^{-1}$ becomes
larger than $\tau _{hs}^{-1}\approx \omega _{c}w_{hs}n_{hs}/2\pi $, one
should observe a usual quadratic MR. On contrary, at lower temperature and
in higher field in the CDW state, when $\tau _{hs}^{-1}>\tau _{i+ph}^{-1}$,
the linear MR should be observed as a general phenomenon. This crossover is
clearly seen on experimental data in Figs. \ref{F1}-\ref{F3}, and, according
to the theoretical model, the crossover field increases with temperature, as
shown in Fig. \ref{F4}.

Let us discuss the possible microscopic origin of the electron scattering in
the hot spots in more detail. The mechanism of linear MR above the CDW
transition temperature $T_{CDW}$, proposed in Ref. \cite{FalikovLinearMR},
assumes a strong scattering by soft phonons with a wave vector close to the
nesting vector $Q_{N}$ due to the Peierls instability. Then the hot spots
are those connected by the CDW wave vector, as shown in Fig. \ref{F5}a.
However, in our experiment the linear MR is observed much below $T_{CDW}$.
Then the expected hot spots are the ends of the ungapped FS parts, as shown
in Fig. \ref{F5}b. In these hot spots the FS reconstruction due to CDW is
the strongest, and the electron dispersion depends strongly on the CDW order
parameter. Therefore, electrons in such hot spots may be easily scattered by
CDW fluctuations. Unfortunately, the available experimental data do not give
detailed informations about the Fermi surface in RTe$_{3}$ compounds in the
CDW state: the ARPES data \cite{Brouet08} do not have sufficient resolution
to determine even the FS topology in the CDW state, while the magnetic
quantum oscillations in the CDW state are complicated by the second CDW
transition in some RTe$_{3}$ materials and by magnetic breakdown. Therefore,
there are possibly other hots spots in the ungapped FS parts, and without
detailed information about the FS in the CDW state we cannot predict the
coefficient in the dependence $\tau _{hs}\propto 1/H$.

There are two types of CDW fluctuations: amplitude and phase fluctuations.
Both may arise, e.g., due to the CDW pinning by crystal defects or local
inhomogeneities. The amplitude CDW defects may strongly scatter the
conducting electrons, e.g., due to the inhomogeneous magnetic breakdown (MB),%
\cite{KartsLTP2014} because the MB probability depends exponentially\cite%
{FalikovLinearMR,KaganovSlutskinReview} on the gap opened in the electron
spectrum due to the CDW. At the ends of the gapped region the gap values
decrease, as explicitly shown by ARPES data \cite{Brouet08}, and the
magnetic breakdown become possible even in low field. Moreover, the
amplitude fluctuations of the CDW gap may lead to spatial variations of the
boundary between gapped and ungapped FS parts. Therefore in Fig. \ref{F5}b
we place the hot spots at the ends of the gapped region. The MB amplitude
depends strongly not only on the gap value, but also on the electron
velocity and dispersion in the MB region,\cite{KaganovSlutskinReview} which
is also affected by the amplitude fluctuations of CDW. The inhomogeneous MB
probability leads to the strong electron scattering in the MB regions,
playing the role of hot spots. Note, that this strong scattering mechanism
may be important not only for transverse but also for the longitudinal MR
and may even lead to the phase inversion of magnetic quantum oscillations 
\cite{KartsLTP2014}. The CDW phase fluctuations mean local variations of the
CDW wave vector, which also change the electron dispersion in our hot spots
and affect the MB probability in addition to the direct scattering in these
hot spots by the CDW periodic potential with inhomogeneous wave vector.
Hence, the CDW fluctuations may indeed lead to the strong electron
scattering in the hot spots of the Fermi surface, and, consequently, to
linear MR.

Our theoretical model is in many aspects similar to that in Ref. \cite{FalikovLinearMR}, but there are some important differences. First, the model in Ref. \cite{FalikovLinearMR} is developed and applied only slightly above the CDW transition temperature $T{CDW}$, while our model can be applied much below $T_{CDW}$. Second, the model in Ref. \cite{FalikovLinearMR} is applied only to the unreconstructed FS, while we apply our model to the strongly reconstructed FS. Third, in Ref. \cite{FalikovLinearMR} the CDW fluctuations have the wave-vector equal to the nesting vector, while in our model any Fermi-surface reconstruction due to CDW leads to scattering by CDW fluctuations, even if there are no ungapped Fermi-surface points connected by the CDW wave vector. Fourth, we propose a temperature-driven crossover between the quadratic and linear magnetoresistance, which cannot be found in the model of Ref. \cite{FalikovLinearMR} where the CDW fluctuations are considered only in the vicinity of transition temperature.

In conclusion we have shown that in the the CDW state of RTe$_{3}$ compounds 
there is a crossover from linear magnetoresistance at low temperature to
usual quadratic magnetoresistance at higher temperature. We propose a
general explanation of this phenomenon as being related to the electron
scattering in the hot spots of Fermi surface due to the spatial fluctuations
or inhomogeneity of the charge-density-wave order parameter.

\acknowledgements

The authors gratefully acknowledge the RFBR-CNRS grant 17-52-150007 \ \ P.G.
acknowledges the financial support of the Ministry of Education and Science
of the Russian Federation in the framework of Increase Competitiveness
Program of MISiS. Theoretical part is supported by the Russian Science
Foundation (grants \# 16-42-01100).


\begin{thebibliography}{99}
\bibitem{Gruner} G. Gruner, \textit{Density Waves in Solids} (Addison 
\textendash{} Wesley, Reading, Massachusetts, 1994).

\bibitem{Monceau12} P. Monceau, Advances in Physics \textbf{61}, 325 (2012).

\bibitem{Abrik} A.A. Abrikosov, \textit{Fundamentals of the theory of metals}%
, (North-Holland, Amsterdam, 1988).

\bibitem{Ziman} J. M. Ziman, \textit{Principles of the Theory of Solids},
(Cambridge Univ. Press, 1972).

\bibitem{Kittel} Ch. Kittel, \textit{Quantum Theory of Solids}, (2nd
Edition, John Wiley \& Sons, New York, 1987)

\bibitem{Kapitza} P.L. Kapitza, Proc. R. Soc. Lond. \textbf{A 123}, 292
(1929).

\bibitem{Abrikosov98} A.A. Abrikosov, Phys. Rev. \textbf{58}, 2788 (1998).

\bibitem{Abrikosov99} A.A. Abrikosov, Phys. Rev. \textbf{60}, 4231 (1999).

\bibitem{Littlewood03} M. M. Parish and P. B. Littlewood, Nature \textbf{426}%
, 162 (2003).

\bibitem{Herring60} C. Herring, J. Appl. Phys. \textbf{31}, 1939 (1960)

\bibitem{Richard87} J. Richard, P. Monceau and M. Renard, Phys. Rev. B 
\textbf{35}, 4533 (1987).

\bibitem{Coleman90} R.V. Coleman, M.P. Everson, Hao-An Lu and A. Johnson,
Phys. Rev. \textbf{41}, 460 (1990).

\bibitem{Lei16} Hechang Lei, Kai Liu, Jun-ichi Yamaura, Sachiko Maki,
Youichi Murakami, Zhong-Yi Lu, and Hideo Hosono, Phys. Rev. B \textbf{93},
121101 (2016).

\bibitem{Naito82} M. Naito1, and S. Tanaka, J. Phys. Soc. Jpn. \textbf{51},
228 (1982).

\bibitem{Chen17} H. X. Chen, Z. L. Li, L.W. Guo, and X. L. Chen, Europen
Phys. Lett. \textbf{117}, 27009 (2017).

\bibitem{Chen17Arx} Hongxiang Chen, Zhilin Li, Xiao Fan, Liwei Guo, Xiaolong
Chen, arXiv:1706.08661 {[}cond-mat.mtrl-sci{]}.

\bibitem{Rotger94} A. R�tger, J. Lehmann, C. Schlenker, J. Dumas, J.
Marcus, Z. S. Teweldemedhin and M. Greenblatt, EuroPhys. Lett. \textbf{25},
23 (1994).

\bibitem{Schlenker} C. Schlenker, \textit{Low-Dimensional Electronic
Properties of Molybdenum Bronzes and Oxides} (Kluwer Academic Publishers,
Springer, 1989).

\bibitem{Ru08} N. Ru, C. L. Condron, G. Y. Margulis, K. Y. Shin, J.
Laverock, S. B. Dugdale, M. F. Toney, and I. R. Fisher, Phys. Rev. B \textbf{%
77}, 035114 (2008).

\bibitem{DiMasi95} E. DiMasi, M. C. Aronson, J. F. Mansfield, B. Foran, and
S. Lee, Phys. Rev. B \textbf{52}, 14516 (1995).

\bibitem{Brouet08} V. Brouet, W. L. Yang, X. J. Zhou, Z. Hussain, R. G.
Moore, R. He, D. H. Lu, Z. X. Shen, J. Laverock, S. B. Dugdale, N. Ru, and
I. R. Fisher, Phys. Rev. B \textbf{77}, 235104 (2008).

\bibitem{Ru06} N. Ru and I. R. Fisher, Phys. Rev. B \textbf{73}, 033101
(2006).

\bibitem{Fang07} A. Fang, N. Ru, I. R. Fisher, and A. Kapitulnik, Phys. Rev.
Lett. \textbf{99}, 046401 (2007).

\bibitem{sinchPRL14} A.A. Sinchenko, P.D. Grigoriev, P. Lejay, and P.
Monceau, Phys. Rev. Lett. \textbf{112}, 036601 (2014).

\bibitem{Lavagnini10R} M. Lavagnini, M. Baldini, A. Sacchetti, D. Di Castro,
B. Delley, R. Monnier, J.-H. Chu, N. Ru, I. R. Fisher, P. Postorino, and L.
Degiorgi, Phys. Rev. B \textbf{81}, 081101(R) (2010).

\bibitem{Yao06} H. Yao, J.A. Robertson, Eun-Ah Kim, and S.A. Kivelson, Phys.
Rev. B \textbf{74}, 245126 (2006).

\bibitem{SinchPRB12} A.A. Sinchenko, P. Lejay, and P. Monceau, Phys. Rev. B 
\textbf{85}, 241104(R) (2012).

\bibitem{Young1968} R.A. Young, Phys. Rev. \textbf{175}, 813 (1968).

\bibitem{FalikovLinearMR} L. M. Falicov and Henrik Smith, Phys. Rev. Lett. 
\textbf{29}, 124 (1972).

\bibitem{Comment1} In quasi-2D metals in strong magnetic field, when the
Landau-level separation $\hbar\omega_c$ exceeds their width $\sim \hbar
/\tau $ and the interlayer transfer integral $t_z$, in addition to quantum
oscillations the effective mean-free time $\tau$ acquires a monotonic
magnetic-field dependence. As a result, at $\hbar\omega_c\gg t_z,\hbar /\tau$
the square-root longitudinal interlayer magnetoresistance appears\cite%
{WIPRB2011,WIPRB2012,GrigPRB2013,Grigbro} due to the effective
remormalization and the square-root field dependence of the imaginary part
of electron self-energy (or of the effective mean scattering rate $1/\tau $).

\bibitem{KartsLTP2014} M.V. Kartsovnik, V.N. Zverev, D. Andres, W.
Biberacher, T. Helm, P.D. Grigoriev, R. Ramazashvili, N.D. Kushch, H.
Muller, Low Temp. Phys. \textbf{40}, 377 (2014).

\bibitem{KaganovSlutskinReview} MI. Kaganov and A.A. Slutskin, Physics
Reports (Review Section of Physics Letters) \textbf{98}, 189-271 (1983) .

\bibitem{WIPRB2011} P.D. Grigoriev, Phys. Rev. B \textbf{83}, 245129 (2011).

\bibitem{WIPRB2012} P. D. Grigoriev, M. V. Kartsovnik, W. Biberacher, Phys.
Rev. B \textbf{86}, 165125 (2012).

\bibitem{GrigPRB2013} P.D. Grigoriev, Phys. Rev. B \textbf{88}, 054415
(2013).

\bibitem{Grigbro} A. D. Grigoriev, P. D. Grigoriev, Low Temp. Phys. \textbf{%
40}, 377 (2014).
\end{thebibliography}
\end{document}